\documentclass[slac_one]{revtex4}
\usepackage{graphicx}
\usepackage{fancyhdr}
\begin{document}

\title{{\small{2005 ALCPG \& ILC Workshops - Snowmass,
U.S.A.}}\\ 
\vspace{12pt}
Modular Implementation of Particle Flow Algorithm with 
Minimized Dependence on the Detector Geometry}

\author{A.Raspereza}
\affiliation{DESY, Hamburg 22607, Germany}

\begin{abstract}
A Particle Flow Algorithm (PFA) 
with the minimized dependence on the detector geometry 
is presented. Current PFA implementation includes procedures 
of the track reconstruction, calorimeter clustering, 
and individual particle reconstruction and is meant as a tool 
for the optimization of the International $e^+e^-$
Linear Collider detector.

\end{abstract}

\maketitle

\thispagestyle{fancy}

\section{INTRODUCTION} 
The most promising strategy for event reconstruction at 
the future linear $e^+e^-$ collider experiment is based 
on the particle flow concept, implying reconstruction of the four-vectors
of all particle produced in an event. The particle flow
algorithm works best at moderate energies of individual 
particles, below about 100 GeV. In this regime, the 
tracking system reconstructs the momentum 
of the charged particles with an accuracy superseding 
the energy and angle measurements
with calorimeters. Hence, in order to attain a better reconstruction of 
events, the charged particle measurement must be solely based 
on the tracking information. The crucial step of the 
particle flow algorithm is correct assignment of calorimeter
hits to the charged particles and efficient separation of 
close-by showers produced by charged and neutral particles.
Monte Carlo studies have shown that an ideal reconstruction
algorithm~\cite{pflow_morgunov}, 
which finds each particle and measures its energy and 
direction with the detector resolution expected for single particles,
could reach a jet energy resolution of 14\%/$\sqrt{E}$.
Over the years a jet energy resolution of 30\%/$\sqrt{E}$ 
has become accepted as a good compromise between the theoretically
possible and practically achievable resolution. 

In this paper modular implementation of the Particle Flow algorithm (PFA) 
with weak dependence on the detector geometry is presented. It is meant as 
a tool for the linear collider detector optimization. 

\section{PARTICLE FLOW ALGORITHM IN MARLIN}
Particle Flow algorithm is implemented in a modular way within the 
framework of the MARLIN package~\cite{Marlin}. Algorithm consists of the following 
steps: 
\begin{itemize}
\item{track finding and fitting in the main tracking device;}
\item{cluster finding in calorimeters;}
\item{track $-$ cluster matching and reconstruction of individual particles.}
\end{itemize}
Each step is implemented as a separate module, MARLIN processor. All processors 
constitute MarlinReco package which can be downloaded from the web~\cite{MarlinReco}.

\subsection{Track Finding and Fitting}
Two separate track finding algorithms are available within the MarlinReco package. 
The first one is based on the existing LEP code 
and optimized for the Time-Projection-Chamber (TPC) as the main tracking device. The algorithm
exploits Kalman filter approach for track finding and fitting, taking into account 
particle interaction with the detector material such as
ionization losses and multiple scattering. 
The second algorithm is designed for silicon 
tracker, which has relatively small number of layers. Algorithm represents 
combinatorial search for set of hits compatible with the helix hypothesis. 

\subsection{Calorimeter Clustering}
Cluster finding in calorimeters is based solely on the spatial information. 
Algorithm requires as an input the list of calorimeter hits with their 
coordinates. No amplitude information is used in the clustering procedure,
making the algorithm applicable to both analogue and digital calorimeters. 
Clustering is applied on the unified array of hits in the electromagnetic
and hadronic calorimeters and produces as an output the list of found clusters. 
Detailed description of the algorithm can be found 
in Reference~\cite{TrackwiseClustering}. 
Found clusters are classified into four-categories on the basis of 
the cluster shape analysis.
\begin{itemize}
\item{The electro-magnetic clusters, whose longitudinal profile is compatible 
with an expectation from electrons or photons.}
\item{The MIP (minimal ionizing particle) clusters, whose shape is 
compatible with the helix model. In addition an energy 
of such clusters is required to be compatible with an expectation from MIP.}  
\item{The hadronic clusters; these are clusters not classified as the MIP or 
electromagnetic clusters.}
\end{itemize}

\subsection{Track$-$Cluster Matching and Individual Particle Reconstruction}
Once tracking and the calorimeter clustering is performed, an attempt is made
to associate clusters with tracks. For each track, its intersection
point with the front face of the electromagnetic calorimeter is determined. 
Cluster containing calorimeter hit closest to this intersection point is 
found. If the distance from the intersection point to the closest hit 
is less than certain predefined threshold, cluster is associated with the track.

Electromagnetic clusters with no associated track are identified as photons,
whereas electromagnetic clusters with associated track are regarded as electrons/positrons.
MIP clusters with associated track are identified as muons. Hadronic clusters with 
associated tracks are accepted as charged pion candidates. 
Finally, hadronic clusters with no associated track are identified as neutral hadrons. 
Four-momentum of charged objects are estimated using tracking information. Track parameters
at the point of closest approach to the primary interaction point define 
momentum vector of charged objects (electrons, muons, charged hadrons).
For neutral objects, cluster energy is used as an estimate of particle energy, while
the line connecting interaction point with the cluster centroid is used as 
an estimate of the direction of particle momentum vector.

\section{RESULTS}

\begin{figure}[h]
\begin{minipage}[c]{0.45\textwidth}
\centering
\includegraphics[width=1.0\textwidth]{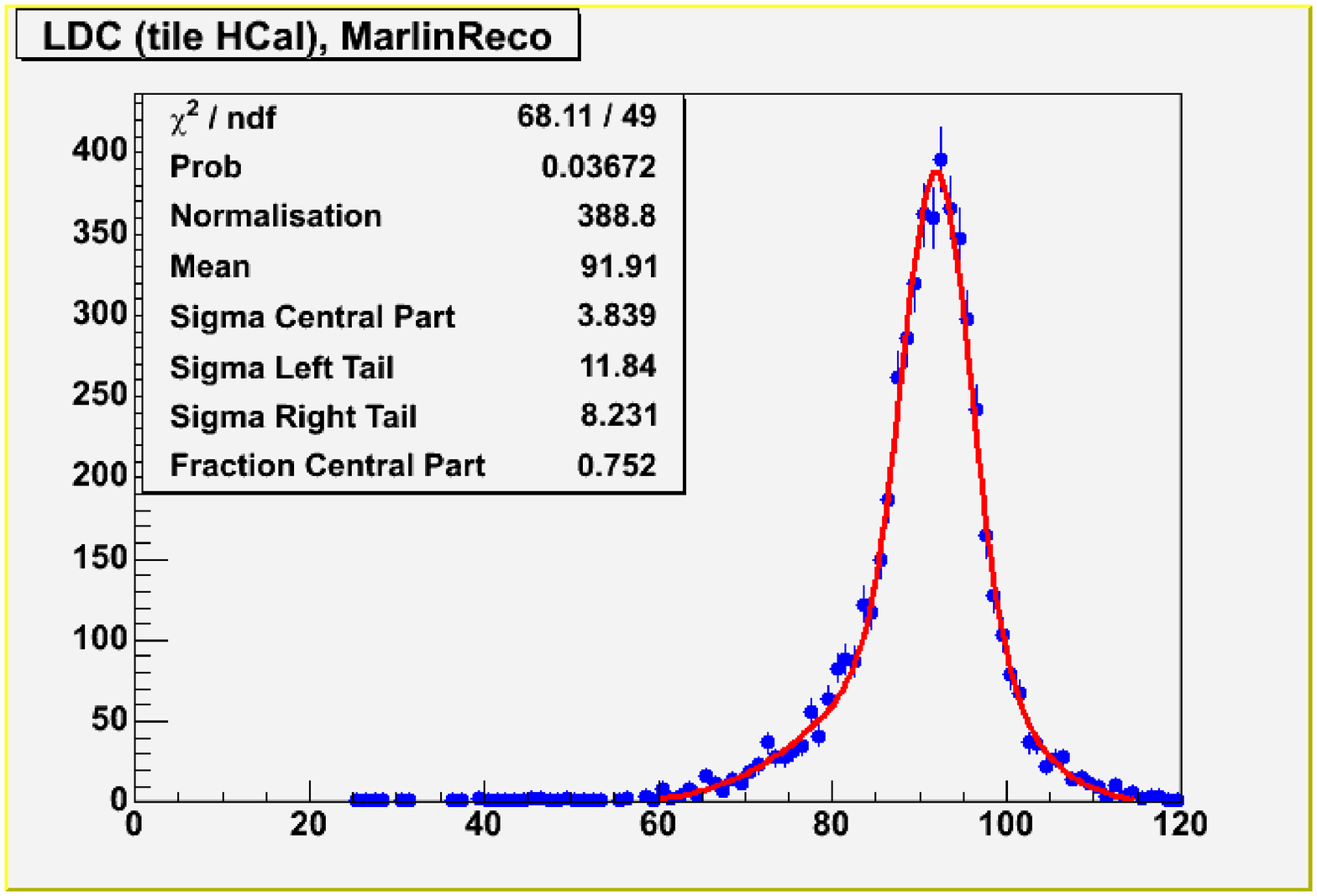}
\caption{Reconstructed visible mass in the sample of 
$Z\rightarrow q\bar{q}(q=u,d,s)$ events at center-of-mass energy of 91.2 GeV 
for the LDC detector with the analogue tile HCAL.
\label{fig:Z0_LDCscint}}
\end{minipage}
\begin{minipage}[c]{0.05\textwidth}
$\phantom{0}$
\end{minipage}
\begin{minipage}[c]{0.45\textwidth}
\centering
\includegraphics[width=1.0\textwidth]{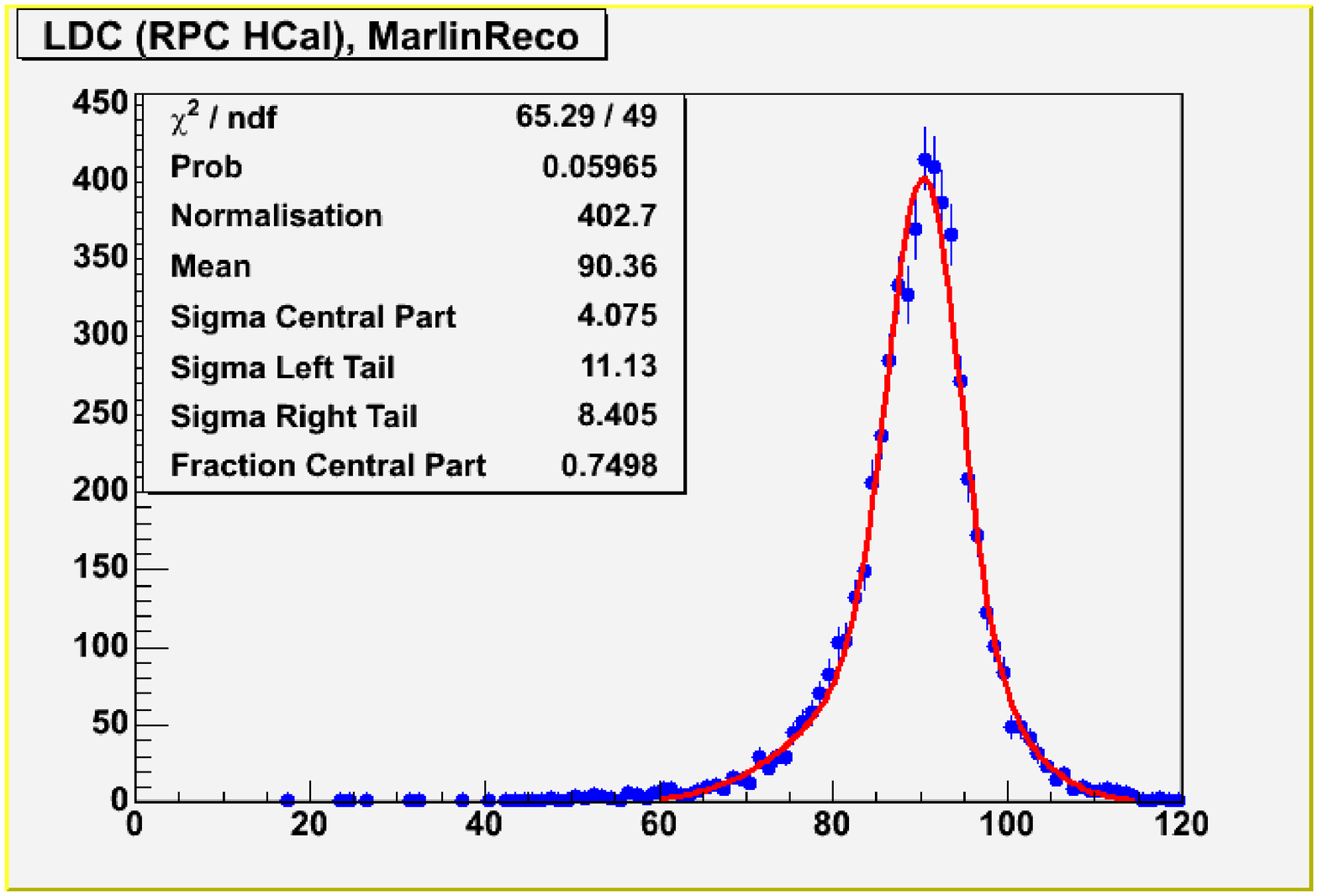}
\caption{Reconstructed visible mass in the sample of 
$Z\rightarrow q\bar{q}(q=u,d,s)$ events at center-of-mass energy of 91.2 GeV 
for the LDC detector with the digital RPC HCAL.  
\label{fig:Z0_LDCrpc}}
\end{minipage}
\end{figure}

\begin{figure}[h]
\begin{minipage}[c]{0.45\textwidth}
\centering
\includegraphics[width=1.0\textwidth]{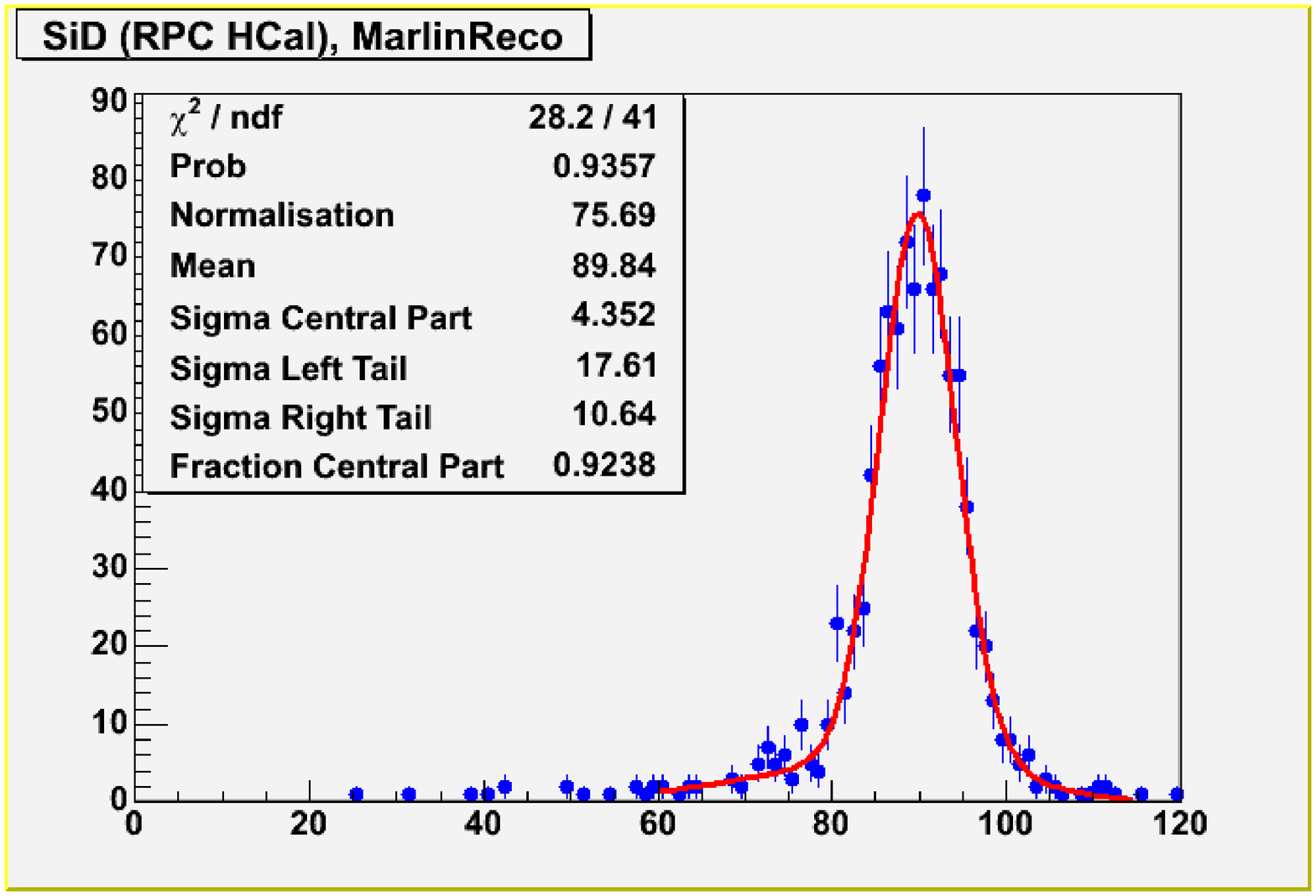}
\caption{Reconstructed visible mass in the sample of 
$Z\rightarrow q\bar{q}(q=u,d,s)$ events at center-of-mass energy of 91.2 GeV 
for the SiD detector with the digital RPC HCAL. 
\label{fig:Z0_SiD}}
\end{minipage}
\begin{minipage}[c]{0.05\textwidth}
$\phantom{0}$
\end{minipage}
\begin{minipage}[c]{0.45\textwidth}
\centering
\includegraphics[width=1.0\textwidth]{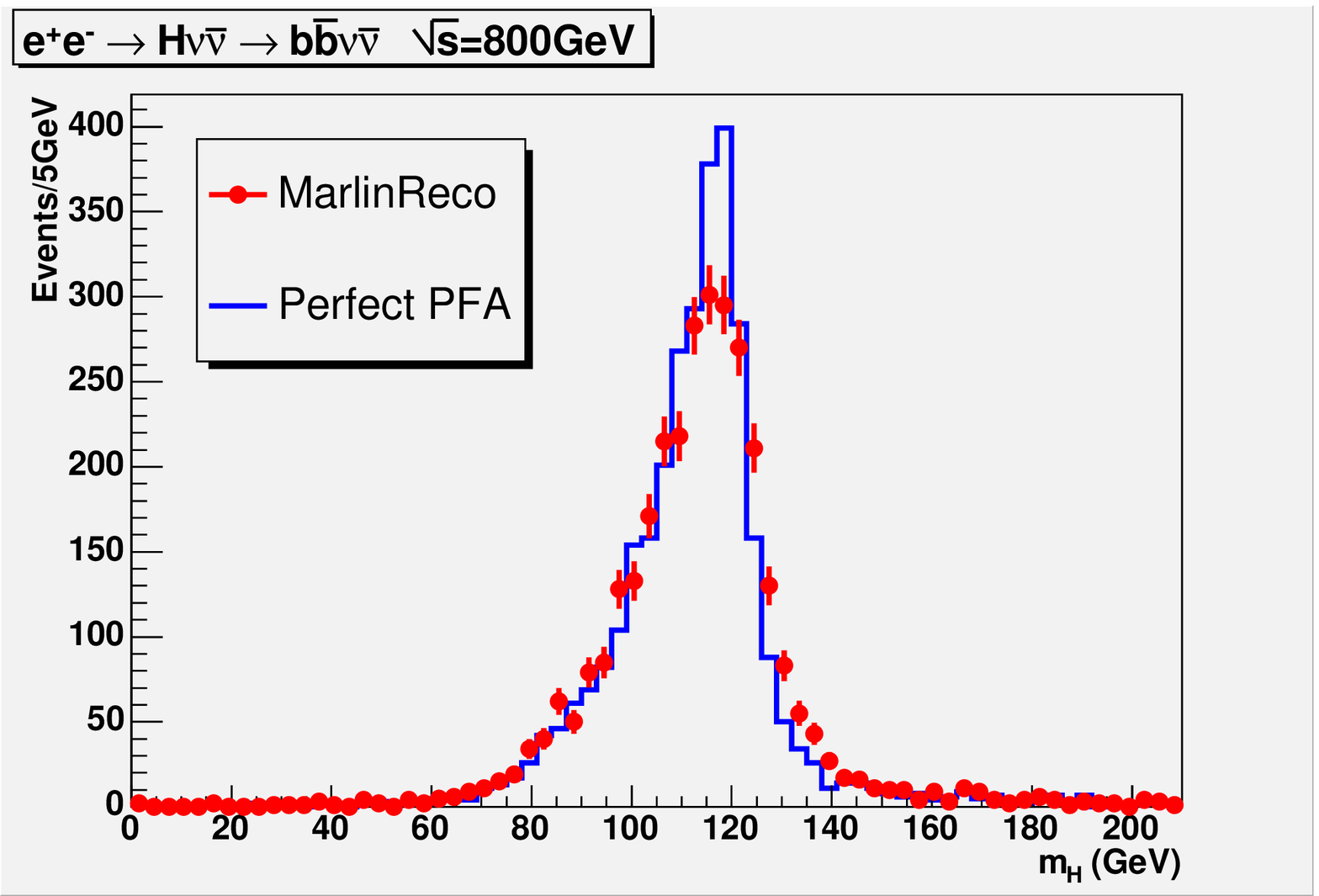}
\caption{Reconstructed Higgs boson mass in the sample of 
$e^+e^-\rightarrow W^+W^-\nu\bar{\nu}\rightarrow H\nu\bar{\nu}$ events 
at center-of-mass energy of 800 GeV for the LDC detector with 
analogue tile HCAL. The simulated Higgs boson mass is 120 GeV.
Result of realistic PFA (dots) is compared with the perfect reconstruction (histogram).
\label{fig:Hvv}}
\end{minipage}
\end{figure}

Performance of the algorithm has been tested with the sample of hadronic events
at Z-pole. The algorithm 
is applied to the different detector models. Figures~\ref{fig:Z0_LDCscint} 
and~\ref{fig:Z0_LDCrpc} present the reconstructed visible mass for the LDC 
(Large Detector Concept) detector with the TPC as the main tracking device. 
Detector simulation is performed with 
the program Mokka~\cite{Mokka}. Figure~\ref{fig:Z0_LDCscint} corresponds to the LDC
detector with the analogue W-Si electromagnetic calorimeter (ECAL) and analogue 
hadron calorimeter consisting of steel absorber plates interleaved with scintillating tiles. 
Figure~\ref{fig:Z0_LDCrpc} corresponds to the LDC detector with the analogue W-Si 
ECAL and digital HCAL consisting of steel absorber plates interleaved with 
the resistive-plate chambers (RPC) as an active elements. Figure~\ref{fig:Z0_SiD} presents
the reconstructed visible mass for the small detector with the silicon tracker (SiD), 
W-Si ECAL and digital RPC HCAL. 
The detector response is simulated with the SLIC program~\cite{SLIC}. 
The resolution achieved varies from 40 to 45\%, depending on the detector model.

Additionally, PFA performance has been tested with the the selected signal processes
at higher center-of-mass energies.
As an example, Figure~\ref{fig:Hvv} presents reconstructed Higgs boson mass for the 
fusion process, $e^+e^-\rightarrow W^+W^-\nu\bar{\nu}\rightarrow H\nu\bar{\nu}$ 
with subsequent Higgs boson decays into b quarks. Process is simulated at center-of-mass
energy of 800 GeV. The Higgs boson mass is 120 GeV. The detector response is simulated 
with Mokka for the LDC detector with the analogue tile HCAL.

\section{FUTURE DEVELOPMENTS}
The algorithm described in this paper is incomplete and needs further
development. We hope that PFA performance can be significantly improved by 
\begin{itemize}
\item{supplementing tracking in the main tracker with the dedicated pattern
recognition in the vertex detector and forward tracking devices (this will 
increase track finding efficiency of low $P_T$ tracks);}
\item{inclusion of the dedicated neutral vertex and kink finding procedures 
in the chain of PFA;}
\item{further optimization and refinement of the clustering algorithm.}
\end{itemize}


\begin{thebibliography}{9} 

\bibitem{pflow_morgunov}
V. Morgunov,  {\it{"Calorimetry Design With Energy Flow Concept (Imaging Detector for High Energy Physics)"}}, 
10th International Conference 
on Calorimetry in High Energy Physics (CALOR 2002), 
Pasadena, California, 25-30 Mar 2002,
published in Pasadena 2002, "Calorimetry in particle physics" 70-84.

\bibitem{Marlin}
F. Gaede,  {\it{"Marlin et al: Introduction to ILC-LDC Simulation and Reconstruction Software"}}, 
ALCPG0806, these proceedings.

\bibitem{MarlinReco}
http://www-zeuthen.desy.de/linear\_collider

\bibitem{TrackwiseClustering}
A. Raspereza {\it{"Clustering in MARLIN, PFlow and Detector Optimization"}}, talk given at CALICE Collaboration Meeting , 
October 2005; talk available at http://www-flc.desy.de/flc/science/hcal/index.html

\bibitem{Mokka}
http://polywww.in2p3.fr/geant4/tesla/www/mokka/mokka.html

\bibitem{SLIC}
J. McCormick, {\it{"Full Detector Simulation using SLIC and LCDD"}}, ALCPG0803, these proceedings.

\end{thebibliography}
\end{document}